\documentclass[manuscript]{aastex}

\newcommand{\kms}{km~s$^{-1}$}
\newcommand{\ro}{R$_{\odot}$}

\slugcomment{Not to appear in Nonlearned J., 45.}

\shorttitle{The coronal green line profiles}

\begin{document}


\title{Analysis of the coronal green line profiles: evidence of excess 
blueshifts}


\author{K. P. Raju}
\affil{Indian Institute of Astrophysics, Bangalore, India}
\email{kpr@iiap.res.in}

\and

\author{T. Chandrasekhar and N. M. Ashok}
\affil{Physical Research Laboratory, Ahmedabad, India}
\email{chandra@prl.res.in}
\email{ashok@prl.res.in}

\begin{abstract}
The coronal green line (Fe XIV 5303 \AA) profiles were obtained from Fabry-Perot 
interferometric observations of the solar corona during the total solar 
eclipse of 21 June 2001 from Lusaka, Zambia. The instrumental width
is about 0.2 {\AA} and the spectral resolution is about 26000. About 300 line
profiles were obtained within a radial
range of 1.0--1.5 {\ro} and position angle coverage of about 240\degr. The 
line profiles were fitted with single Gaussian and their 
intensity, Doppler velocity, and line width have been obtained. Also 
obtained are the centroids of the line profiles which give a measure of line
asymmetry. The histograms of Doppler velocity show excess blueshifts while the
centroids reveal a pre-dominant blue wing in the line profiles.
It has been found that the centroids and the Doppler velocities are
highly correlated. This points to the presence of multiple
components in the line profiles with an excess of blueshifted components. 
We have then obtained the(Blue--Red) wing intensity which clearly reveals 
the second component, majority of which are blueshifted ones. This confirms 
that the coronal green line profiles often contain multicomponents with excess
blueshifts which also depend on the solar activity. The magnitude of the Doppler 
velocity of the secondary component is in the range 20--40 {\kms} and they show 
an increase towards poles. The possible explanations of the multicomponents could be 
the type II spicules which were recently found to have important to the coronal heating
or the nascent solar wind flow, but the cause of the blue asymmetry in the coronal lines
above the limb remains unclear. 
\end{abstract}

\keywords{Sun: corona --- Sun: transition region}

\section{Introduction}

The coronal green line Fe XIV 5302.86 {\AA} is the most prominent visible 
emission line from the inner solar corona and hence it is the most widely 
observed line during the total solar eclipses. This is because it's formation 
temperature is about 1.8 MK which is closer to the average temperature of the 
inner corona.

Line profile analysis gives information on the physical conditions
of the source such as density, temperature, Doppler and nonthermal velocities,
wave motions  etc. Such analysis on the coronal green line can provide useful 
insights to the unresolved problems such as the coronal heating and the 
acceleration of solar wind. 

The existence of mass motions in the corona remained controversial in the past.
The inner corona was thought to be quiescent with no mass motions greater than
a few \kms \citep{Newk67,Lieb75,Singh82}. However, there have been several 
observations of large-scale motions in the corona 
\citep{Delone69,Delone75,Delone88,Chandra91,Raju93}. The later SOHO and TRACE
observations have laid to rest this controversy by showing a highly dynamic
corona with large velocities and different kinds of wave motions \citep{Brek99}. 

It has been known since seventies that the EUV emission lines from the 
transition region of the quiet Sun are systematically redshifted \citep{Brek99}.
The typical value of the average downflow velocity is 5--10 {\kms}. The 
magnitude of the redshift has been found to increase with temperature
and then decrease sharply \citep{Doschek76,Hasl91,Brek93}. 
Measurements of this variation is somewhat ambiguous but the recent SUMER
results suggest that the upper
transition region and lower corona lines are blueshifted, with a steep 
transition from red to blue shifts above 0.5 MK \citep{Chae98,Peter99}.
In active regions, multiple flows were observed by \citet{Kjel88,Kjel93}.
\citet{Brek92} obtains multiple Gaussian fits to Si IV 1402 {\AA} 
profiles with velocities up to 105 {\kms}. \citet{Brek97} reports the
first observation of large Doppler shifts in individual active region loops
above the limb. The high shifts are present only in parts of loops. The 
line of sight velocities are -60 {\kms} and 50 {\kms}, so the axial flow 
velocities could be much higher. A systematic investigation by 
\citet{Kjel98} confirmed that high Doppler shifts are common in 
active region loops. From SUMER observations, \citet{Peter01} finds that 
the emission line profiles of the transition region are best-fitted by a double 
Gaussian with a narrow line core and a broad second component.

The details of line profiles from the corona were difficult to obtain during
the SOHO era and before because of instrumental limitations. The spectral 
resolution of Extreme ultra violet Imaging Spectrometer  
(EIS; \citet{Culhane07}) onboard Hindode \citep{Kosugi07} is about 4000 
which is worse than SUMER but the good signal-to-noise ratio provided the 
possibility of studying the details of the line profile \citep[see][]{Peter10}. 
\citet{Hara08} observed asymmetry in the coronal line profiles of 
Fe XIV 274 {\AA} and
Fe XV 284 {\AA} using EIS/HINODE. The excess emission seen in the blue wing 
has been interpreted in terms of nanoflare heating model by \citet{Pats06}.
\citet{Depont09} find a strongly blueshifted component in the coronal 
emission
lines which is interpreted as due to type II spicules. Blueshifts of about 
30 {\kms} have been found in coronal lines for plasma in the coronal hole
which were interpreted as evidence for nascent solar wind flow \citep{Tu05,Tian10}.
\citet{Peter10} examined line profiles of Fe XV from 
EIS onboard Hinode and found that the spectra are best fit by a narrow line
core and a broad minor component with blueshifts up to 50 {\kms} .
\citet{Depont11} have used SDO and Hinode observations to reveal ubiquitous
coronal mass supply due to acceleration of type II spicules into the corona,
which plays a substantial role in coronal heating and energy balance.

As mentioned above, there have been occasional reports of high velocities in 
the corona even in the pre-SOHO era. These are ground-based eclipse or 
coronagraphic observations in the visible emission lines made above the limb. 
The presence of multicomponents with an excess of blueshifts in the coronal
green line profiles have been reported \citep{Raju93,Raju99}. Recently 
\citet{Tyagun10} reported similar results in the coronal red line
Fe X 6374 {\AA}.
In the present paper, we revisit the problem of the velocity field in the 
corona in the light of new results from SOHO, Hinode etc. We have obtained 
the coronal green line profiles from Fabry-Perot
interferometric observations during the total solar eclipse which have high 
spectral resolution. In the following sections we describe the data and the
analysis steps, results, discussion and conclusions.

\section{Data and Analysis}

Fabry-Perot interferometric observations of the solar corona were made during
the total solar eclipse of 21 June 2001 from Lusaka, Zambia. The instrumental
setup is similar to the earlier observations \citep{Chandra84}. The free spectral
range of the Fabry-Perot interferometer is 4.75 {\AA} and the instrumental 
width is about 0.2 {\AA}. The spectral resolution at the coronal green line is 
about 26000. 


The analysis involved the following steps; i) locating the fringe center 
position in the interferogram, ii) radial scans from the fringe center and 
isolation of fringes, iii) positional identification in the corona, iv) 
wavelength calibration, v) continuum subtraction, vi) Gaussian fitting
to the line profile which gives intensity, linewidth, Doppler velocity. 

The centroid of the line profile which is defined as the wavelength point
that divides the area of the line profile into two was also obtained.
This gives a measure of the line asymmetry if multiple components are present.

About 300 line profiles within a radial range of 1.0--1.5 \ro and position
angle coverage of about 240\degr have been obtained. Only those with a 
signal-to-noise ratio above 15 were considered in the analysis which limited
their number to 272. Those line profiles with a signal-to-noise ratio less than 
15 were found to have a larger uncertainty in the background intensity. This in 
turn affect the accuracy of the Gaussian fitting of the line profiles. 
The position of the line profiles are marked on an EIT image of the Sun in 
Figure 1 where the north is up and the west is towards the right. The position
angle 0\degr corresponds to the west while 90\degr denotes the north pole. 
The data are scattered because they represent the fringe maxima positions. There
is a also gap of about 120\degr around the south pole. 

\section{Results and Discussion}

In Figure 2, we have shown 30 line profiles alongwith their single Gaussian fits.
The estimated errors in the fitting are about 5 \% in intensity, 2 {\kms}  in 
velocity and  0.03 {\AA} in width. The line profiles do not show explicit evidence
of multicomponents and the fits are generally seem to be satisfactory.

The radial variations of linewidth, Doppler velocity and centroid of all the 
line profiles are given in Figure 3. The straight line fits to the radial 
variations do not show any specific trend. However the variations show
a wave-like pattern sometimes. The absence of trend in the width
does not agree with some of the earlier results. For example, \citet{Singh06} 
find that the width
of the green line decreases with coronal height up to about 1.31 Ro and then 
remain constant. The coronal red line showed an opposite behavior. \citet{Chandra91} 
find a broad peak in the radial variation of the width at about 1.2 \ro.

Next we examine the position angle dependence of the radial variations of the 
Doppler velocity and width. The behavior of two position angle intervals 
(155--175, -(45--25)) is shown in Figure 4. The other position angle intervals show
an intermediate behavior. There is an increase of Doppler 
velocity and width with respect to the coronal radial distance in the former 
whereas there is no such dependence in the latter. The wave-like appearance is more 
prominent in position angle intervals. The result imply that the trend is governed by
the underlying activity of the solar region. The increase of the 
Doppler velocity and width with respect to the  coronal radial distance in the
first case implies that there is a dependence of Doppler velocity and width 
which is shown in the lowest panel of Figure 4. The correlation coefficient is 
rather small(0.29) but significant. The probability that any two random distribution 
can give a higher correlation coefficient is only 0.03. A positive correlation between 
Doppler velocity and width could indicate a heating process driving a flow
\citep{Peter10}.

The histograms of width, Doppler velocity and centroid are shown in Figure 5.
The width peaks at 0.9 {\AA} which if it is converted to temperature, is 3.14 MK.
Taking the formation temperature of the line to be 1.8 MK, this would imply a 
nonthermal velocity of 20 {\kms}. The nonthermal velocities in the corona are 
reported to be in the range of 10--100 {\kms} which include the possible 
variations in different coronal regions \citep{Harra99}. The observed average 
nonthermal velocity of agrees well with reported values. The 
histograms of Doppler velocity and centroid are similar. Note that both the 
histograms show a clear excess of blueshifts.

The Doppler velocity versus centroid is given in Figure 6. It can be seen 
that there is a very strong positive correlation between the two ($q>0.99$).
This suggests that there is a secondary component present in the 
line profile because if it is not the case, then the relationship between the
velocity and the centroid will be random. Also the straight-line relationship
seen in both the negative and positive quadrants means that there are both 
blueshifts and redshifts present in the line profile. Hence the results, 
in general, point to the multicomponent-nature of the coronal green
line profiles. 

In order to see the nature of the secondary component, we have obtained the 
(Blue--Red) wing intensity of the line profiles which is plotted in Figure 7.
Only the blue-wing is shown in the Figure. This will bring out 
the secondary component in the line profiles - the positive component represents
the blueshift whereas the negative one gives the redshift. The statistics is 
given in Table 1. Clearly there is an excess of blueshifts over redshifts. Also, for
most of the line profiles in column 1, it is difficult to decide whether they are 
single or multiple because the secondary component is weak which are then put 
together as single/ambiguous.
A Gaussian fit to the secondary component is also shown in the Figure which 
gives the relative intensity, Doppler velocity and the width. Details of the 
fitting procedure are given in Table 2. The relative intensity of the secondary 
component could be up to 54 \%, Doppler velocity $\pm$ (20--40) \kms, and width
0.5--0.8 \AA.

It can be seen that there is a consistent picture emerges from Figures 5--7 and 
Tables 1--2. The histograms of width and centroid in Figure 5 show that there is 
an excess of blueshifts and a prominent blue asymmetry in the line profiles. 
The high positive correlation between the Doppler velocity and centroid in 
Figure 6 implies the presence of multiple components within the line profiles. 
Figure 7 confirms that the multicomponents are real and not any artifacts of the
fitting procedure. Also Table 1 confirms that the prominent blue asymmetry arises
because of the excess blueshifts in the line profiles. This can be further seen in
Table 2. When the single Gaussian fit gives indication of (negative/positive) Doppler
velocity, there is a (blue/red) secondary component  present in the line profile.
This would imply that the parameters obtained through the single Gaussian fitting 
of the line profiles are only indications of the actual values. The Doppler velocities
obtained could be underestimates of the relative line-of-sight velocities between 
the main and secondary components. Similarly the obtained widths may be overestimates
of the individual widths of the components.

The evidence of blue asymmetry in the coronal line profiles was first pointed out by
\citep{Raju93}. The multicomponents were explained on the basis of mass motions in the 
coronal loops\citep{Raju99}, though the blue asymmetry remained as a puzzle.
The occurrence of multicomponents in the coronal line profiles was found to depend upon 
the solar activity. The line profiles of 1980 solar maximum corona (monthly sunspot 
number = 155) showed strong multicomponents, 
sometimes up to 4. The relative velocities between multicomponents were found to go 
up to 70 {\kms}. On the other hand the line profiles of 1983 corona which 
belong to a declining solar activity phase (monthly sunspot number = 91), showed mostly
single Gaussians but
sometimes double Gaussians and rarely triple Gaussians \citep{Chandra91}.
The year 2001 belongs to the solar maximum phase but the activity was lower 
(monthly sunspot number = 134) as compared to 1980. Here the line profiles are  mostly
double Gaussians 
and relative velocities are about 30 {\kms}. Similar results are also observed
for coronal red line, Fe X 6374 {\AA}. From a single Gaussian analysis of 
Norikura coronagraph data, \citet{Raju00} find that though the 
majority of Doppler velocities are only a few {\kms}, there is a definite 
excess of blueshifts over redshifts. Also, \citet{Tyagun10} from an analysis of about
5500 line profiles belong to 1968-72 reported that 80 \% of the coronal red line 
profiles are asymmetric and the fractions of the asymmetric profiles
with more intense blue and red wings are 52 and 28 \% respectively. 
To summarize, the line profiles of coronal visible emission lines often show 
multicomponents with a predominant blue wing. The occurrence of multicomponents
show a dependence on solar activity.

Now what are the possible causes of the blue asymmetry.
The differential rotation of the Sun can cause preferential blueshifts at the east 
limb and redshifts at the west limb. But the maximum velocity is only 2 \kms which is  
comparable to the error involved and hence may not be detected.
The recent results from SOHO and HINODE show evidence of multiple components 
and a predominant blue wing in the EUV emission lines\citet{Hara08,Depont09,Peter10}.
\citet{Depont11} have explained this on the basis of upflows due to type II spicules
which have implications to coronal heating. It is possible that the 
secondary component with a preferential blueshift could be due to the type II spicules. 
The blueshifts are also explained as due to 
the nascent solar wind flow \citep{Tu05,Tian10}. It should be noted that the above 
observations are mostly
made on the disk. The interpretation of the off-limb results is even more complicated due
to the line-of-sight effects. The upflows can cause blueshift, redshift or no shift in 
a line profile depending on the angle between the flow and the plane of the sky. It may 
be noted that \citet{Hara08} find excess blueshifts at the disk center which gradually
disappear near the limb. \citet{Tian10} explain the redshifts of Fe XII and Fe XIII lines
at the limb on the basis of tilt of the solar rotation axis (B0). Our observations are 
made on June 21 when B0 = 1.8. If the flows are assumed to be radial, this may cause a 
slight preferential blueshift at the north pole and redshift at the south pole. However 
we have not seen any preferential blue/redshifts seen at the poles. We have examined the 
dependence of Doppler velocity of the secondary component on the position angle 
which is shown in Figure 8. It may be seen that the velocities show a dependence on
the position angle with a maxima at the poles and minima at the equator. This is 
akin to the behavior of the solar wind flow. It is well-known that the fast wind emanates
from the coronal holes of polar regions whereas the slow wind comes from the streamer
structures in the equatorial regions. Also it is suggested that an asymmetric velocity
distribution of the emitting ions could cause line asymmetries \citep{Peter10}.
It may be seen that the results fit well with the overall behavior of the 
solar atmosphere; that the redshifts in the lower transition region slowly 
changes sign to blueshifts in the upper transition region which continues to 
increase in the lower corona.

The above discussion points to the fact that the velocity field in the corona is 
quite complex. There are evidences of wave motions, mass motions in coronal loops,
solar wind flow, and type II spicular flows. Their detailed nature and significance
are yet to be understood. We may expect that HINODE observations of line profiles from 
both the disk and the limb will give more insights on this.

\section{Conclusions}
\label{S-Conclusions} 

It has been shown that the coronal green line profiles, in general, contains
multicomponents. Though the single Gaussian fitting gives a definite indication of the
of multicomponents, the parameters obtained such as the Doppler velocity and the width 
could be under/over-estimates of the actual values.
The occurrence of multicomponents has been found to be related the solar 
activity. It has also been found that there is a definite blue asymmetry 
meaning an excess of blueshifts over redshifts in the coronal line profiles. 
The causes of the blue asymmetry are not clear but future HINODE observations
of both disk and limb may resolve this.

\acknowledgments

\clearpage

\begin{figure}
\epsscale{1.0}
\plotone{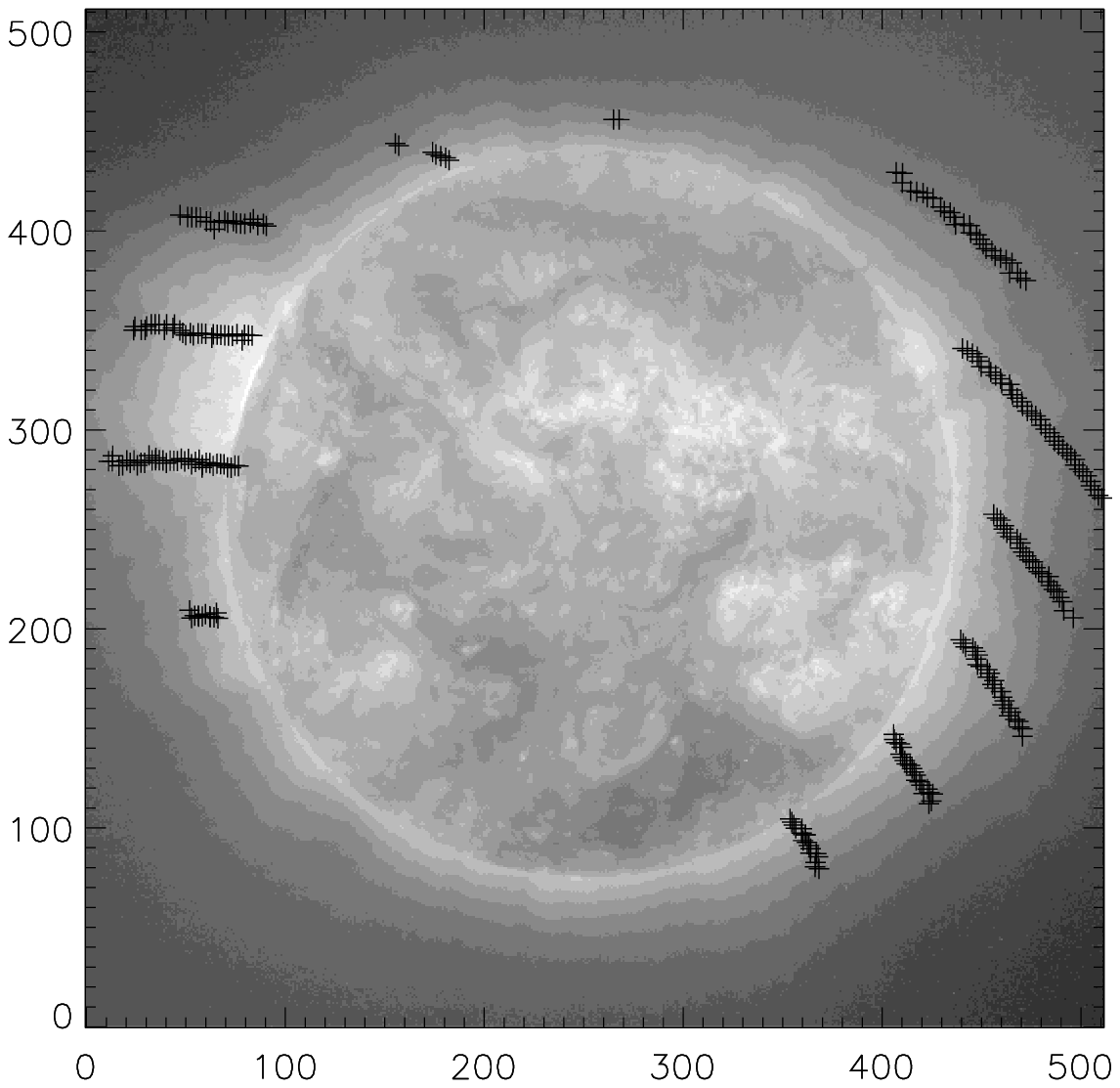}
\caption{The spatial locations of the line profiles marked on an EIT image of the Sun. 
\label{fig1}}
\end{figure}

\clearpage

\begin{figure}
\includegraphics[height=7in,width=6.5in]{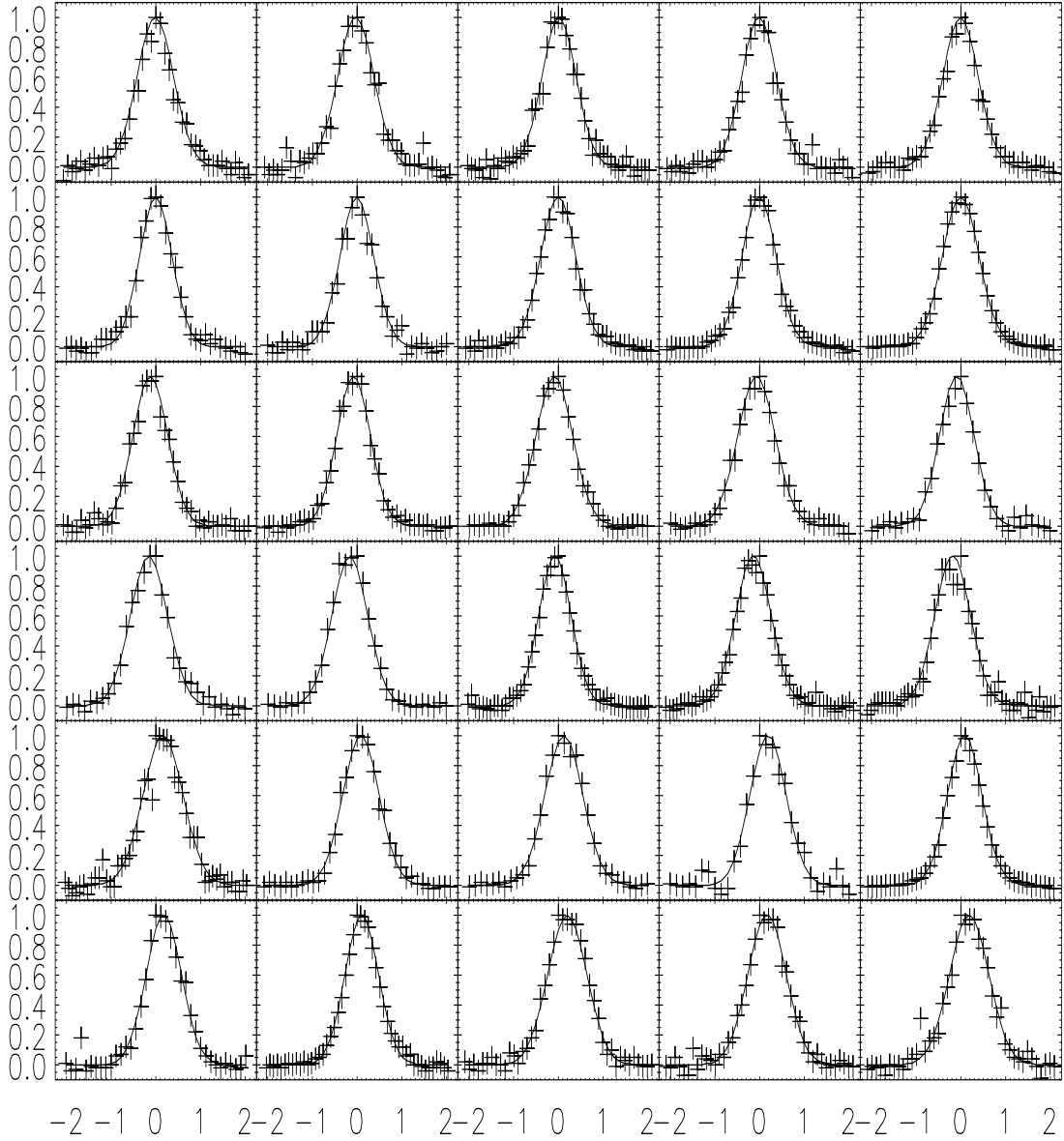}
\caption{Intensity (arbitrary units) plotted against 
wavelength difference from peak (\AA) for 30 line profiles. Single Gaussian 
fittings are given as the continuous line. The fitted parameters are given 
in Table 2. \label{fig2}}
\end{figure}

\clearpage

\begin{figure}
\includegraphics[height=8in,width=6in]{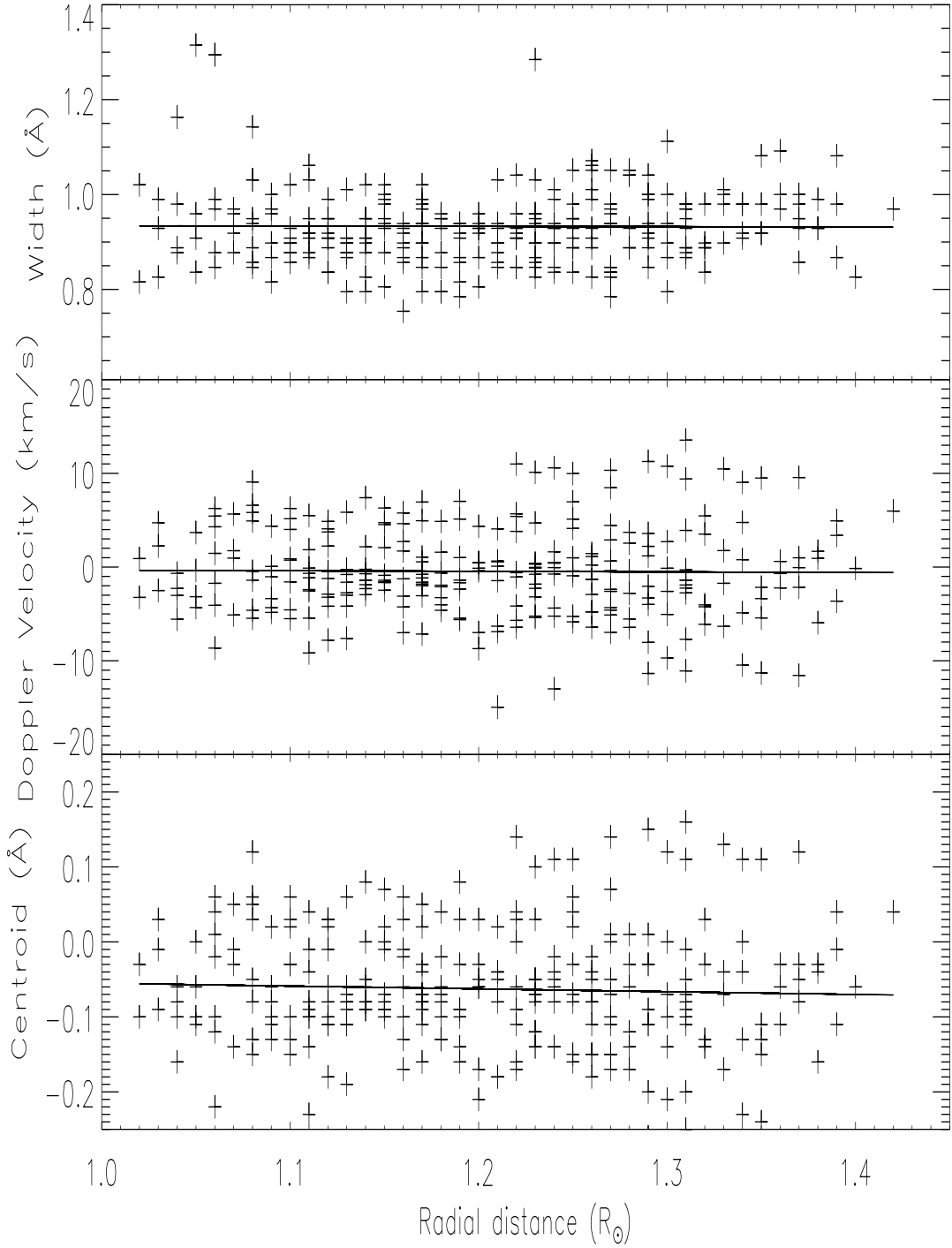}
\caption{Radial variations of linewidth, Doppler 
velocity and centroid. The solid line represents the straight line fit. \label{fig3}}
\end{figure}

\clearpage

\begin{figure}
\epsscale{1.0}
\plotone{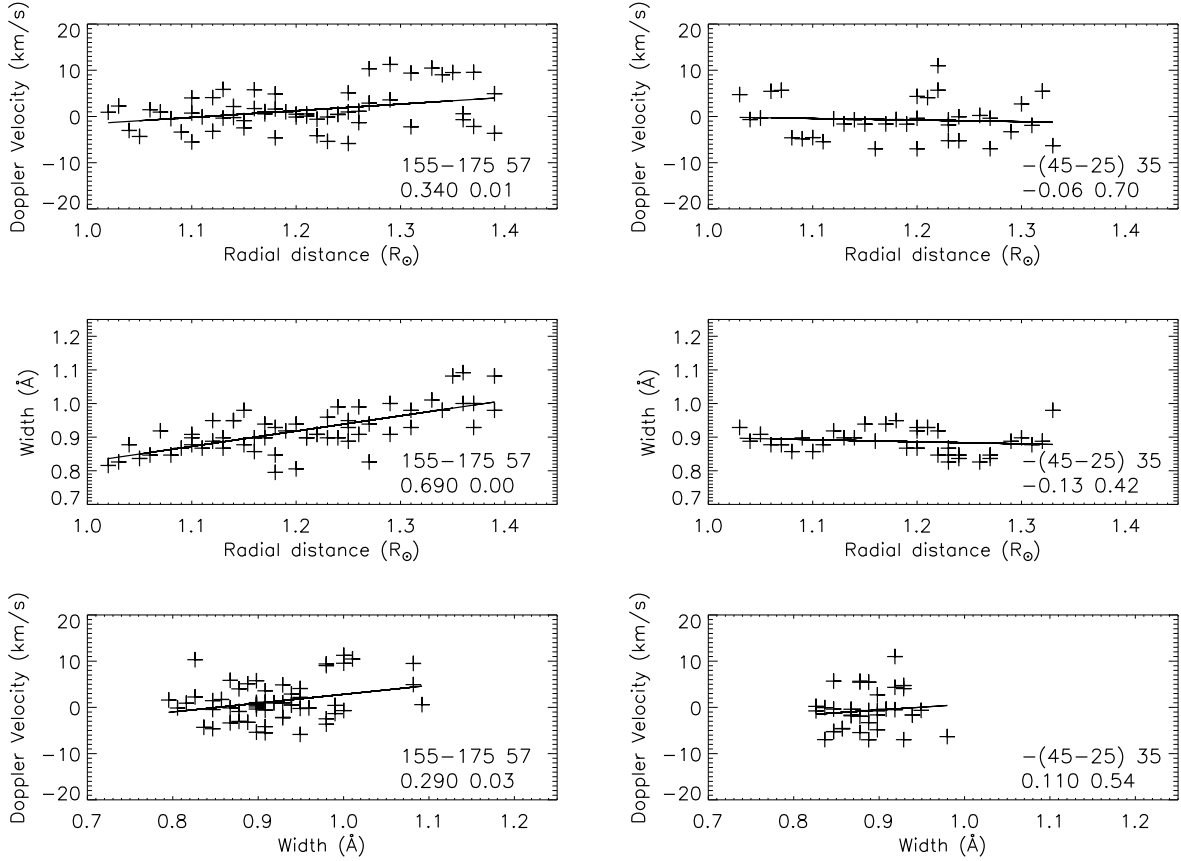}
\caption{Doppler velocity and width are plotted against coronal radial distance for 
two position angles in the two upper panels. Doppler velocity
vs width is plotted in the lowest panel. The solid line represents the 
straight line fit. The numbers indicate the position angle interval, number of
points, correlation coefficient and the probability that any two random 
distribution can give a higher correlation coefficient. \label{fig4}}
\end{figure}

\clearpage
\begin{figure}
\epsscale{1.0}
\plotone{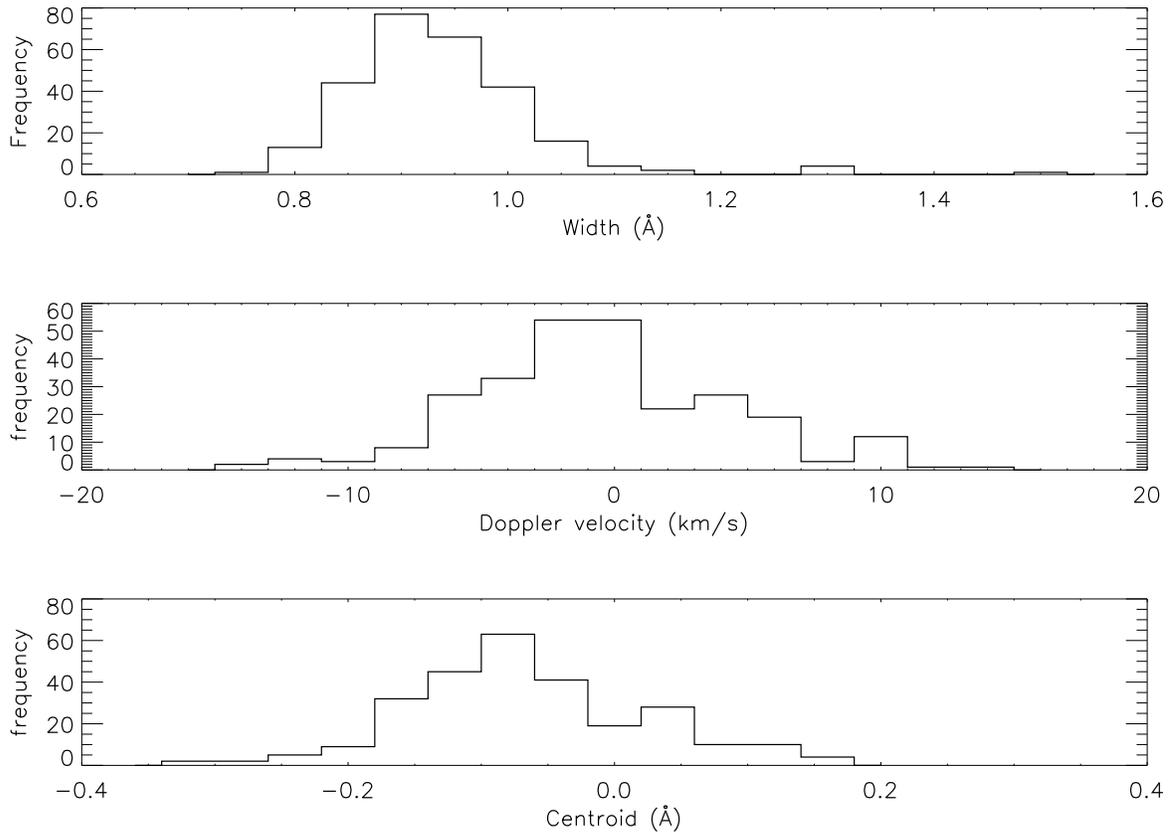}
\caption{Histograms of width, Doppler velocity and centroid. \label{fig5}}
\end{figure}

\clearpage
\begin{figure}
\epsscale{0.7}
\plotone{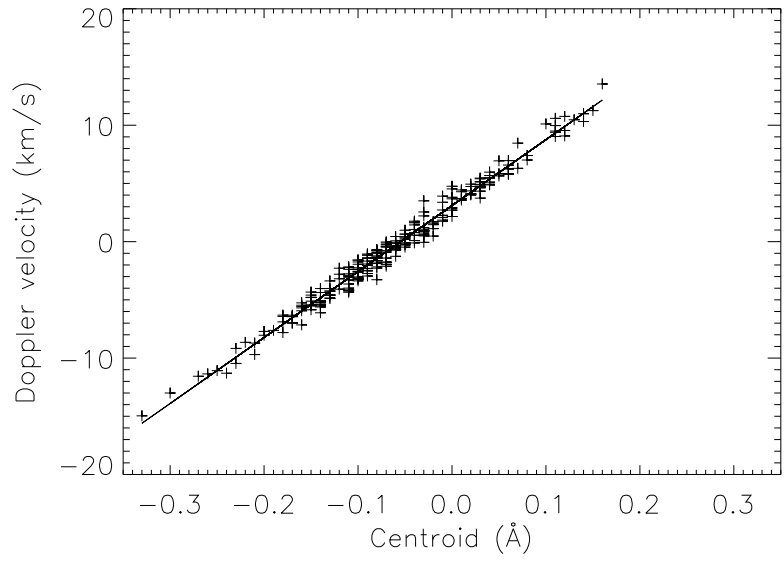}
\caption{Doppler velocity plotted against centroid. The solid line 
represents the straight line fit. \label{fig6}}
\end{figure}

\clearpage
\begin{figure}
\includegraphics[height=7in,width=6.5in]{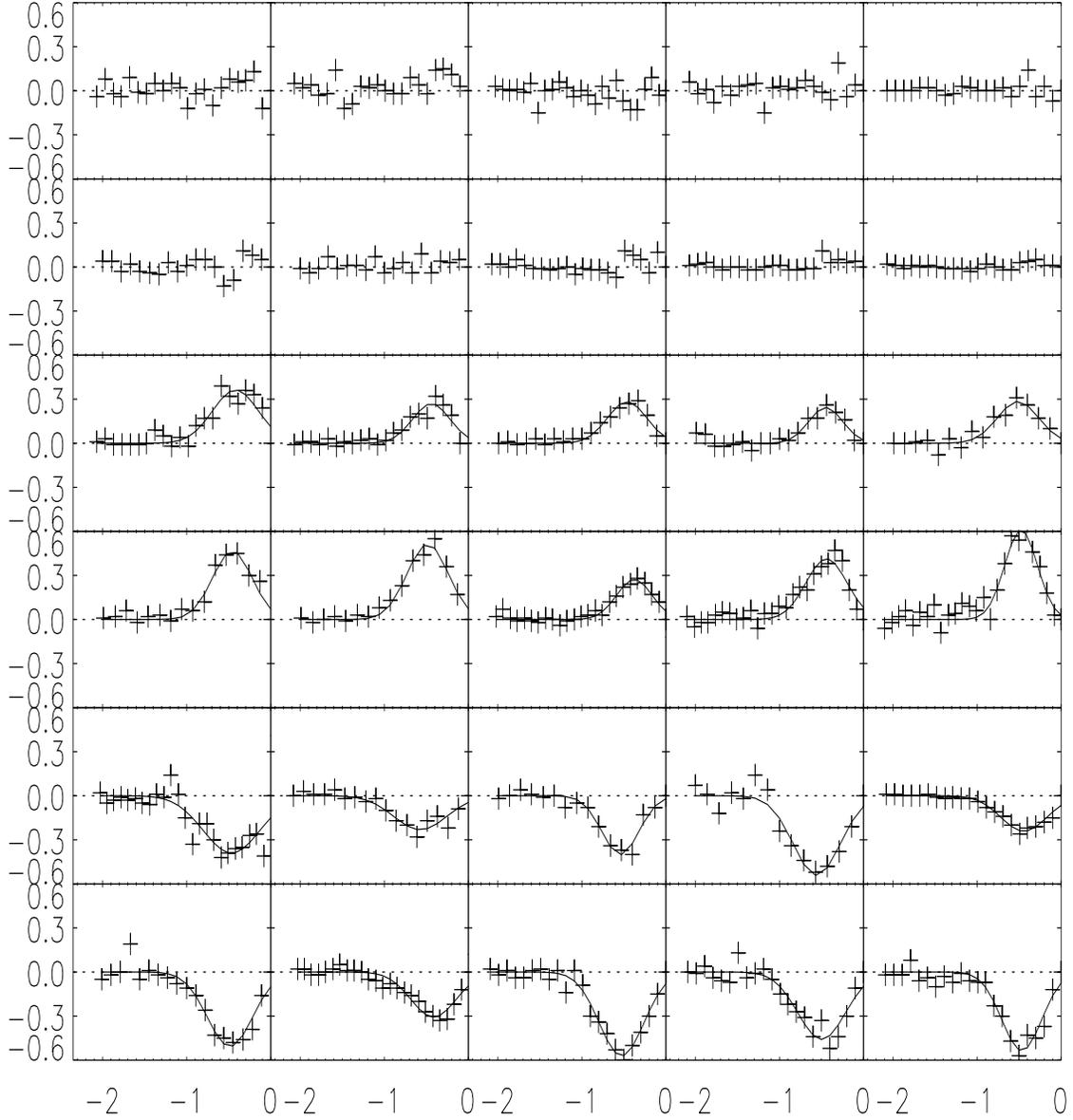}
\caption{The (Blue--Red) wing intensity of the same line 
profiles as of Figure 2. The solid line represents the Gaussian fit. The
fitted parameters are given in Table 2. The upper 10 profiles represent 
single/Ambiguous, the middle 10 represent blueshifted components and the 
lower 10 represent redshifted components.  \label{fig7}}
\end{figure}

\clearpage
\begin{figure}
\epsscale{1.0}
\plotone{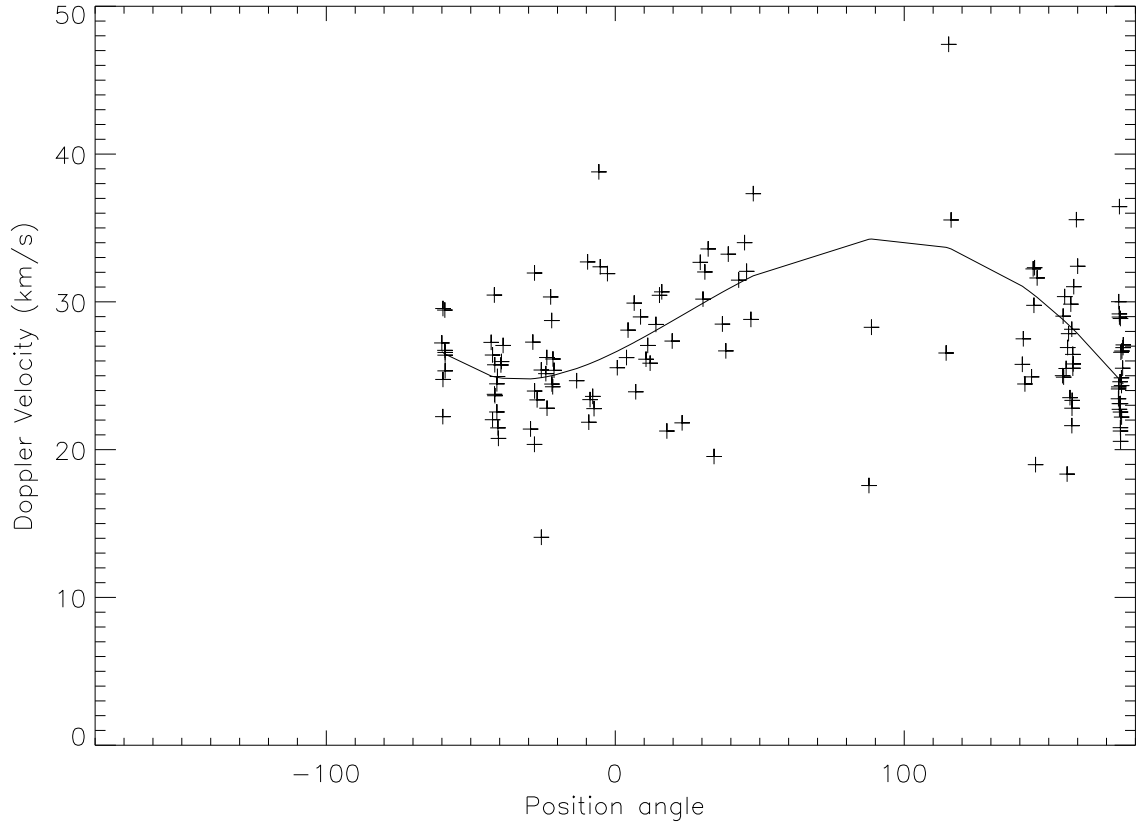}
\caption{Doppler velocity of the secondary component plotted against position 
angle. The solid line represents a polynomial fit. \label{fig8}}
\end{figure}

\clearpage

\begin{deluxetable}{cccc} 
\tablecolumns{4} 
\tablewidth{0pc} 
\tablecaption{Number of single/multicomponents. The numbers given in the bracket 
give the percentage.}
\tablehead{ 
\colhead{Single/Ambiguous} & \colhead{Blue} & \colhead{Red} & \colhead{Total}}
\startdata 
114 & 93 & 65 & 272\\
(42)& (34)&(24)&(100)\\
\enddata
\end{deluxetable}

\clearpage

\begin{deluxetable}{rrrrrrrrrr} 
\tablecolumns{10} 
\tablewidth{0pc} 
\tablecaption{Details of Gaussian fitting. Columns 2--4 give the parameters of 
single Gaussian fitting. Columns 5--7 give the parameters of Gaussian fitting
of the blue component shown in 3--4 rows of Figure 6. Columns 8--10 give the 
parameters of Gaussian fitting of the red component shown in last 2 rows of 
Figure 7.}
\tablehead{ 
\colhead{}    &  \multicolumn{3}{c}{Single Gaussian} &  
 \multicolumn{3}{c}{Blue} & \multicolumn{3}{c}{Red} \\
\cline{2-4} \cline{5-7} \cline{8-10} \\ 
\colhead{No.} & \colhead{Int} & \colhead{Vel} & \colhead{Wid} & \colhead{Int} &
\colhead{Vel} & \colhead{Wid} & \colhead{Int} & \colhead{Vel} & \colhead{Wid}}
\startdata 
   1 &     1.00 &    -0.65 &      0.95&&&&&&\\
   2 &     1.00 &    -1.62 &      0.94&&&&&&\\
   3 &     1.00 &     0.88 &      0.88&&&&&&\\
   4 &     1.00 &     0.72 &      0.90&&&&&&\\
   5 &     1.00 &    -0.49 &      0.92&&&&&&\\
   6 &     1.00 &    -0.06 &      0.84&&&&&&\\
   7 &     1.00 &    -0.40 &      0.87&&&&&&\\
   8 &     1.00 &    -0.37 &      0.91&&&&&&\\
   9 &     1.00 &    -0.67 &      0.89&&&&&&\\
  10 &     1.00 &    -0.24 &      0.95&&&&&&\\
  11 &     1.00 &    -7.00 &      0.89&0.36 &   -23.75 &     0.65&&&\\
  12 &     1.00 &    -4.57 &      0.86&0.27 &   -24.95 &     0.53&&&\\
  13 &     1.00 &    -5.11 &      0.96&0.28 &   -26.13 &     0.55&&&\\
  14 &     1.00 &    -4.08 &     0.97&0.24  &  -25.37  &     0.48&&&\\
  15 &     1.00 &    -5.26 &      0.89&0.29 &   -28.98 &     0.58&&&\\
  16 &     1.00 &    -8.71 &      0.96&0.47 &   -25.85 &     0.55&&&\\
  17 &     1.00 &    -9.17 &      0.92&0.52 &   -27.35 &     0.58&&&\\
  18 &     1.00 &    -4.32 &     0.84&0.27  &  -21.26  &     0.50&&&\\
  19 &     1.00 &    -7.81 &      0.91&0.41 &   -24.86 &     0.57&&&\\
  20 &     1.00 &    -10.46 &     0.92&0.61 &   -26.90 &     0.46&&&\\
  21 &     1.00 &     9.08 &     1.03&&&&-0.39 &   -26.39     &  0.79\\
  22 &     1.00 &     4.74 &     0.93&&&&-0.23 &   -32.37    &  0.74\\
  23 &     1.00 &     6.94 &     0.98&&&&-0.40 &   -30.68 &  0.52\\
  24 &     1.00 &    10.59 &     0.93&&&&-0.54 &   -31.47 &  0.65\\
  25 &     1.00 &     4.92 &     0.89&&&&-0.24 &   -25.00 &  0.65\\
  26 &     1.00 &     9.97 &     0.91&&&&-0.51 &   -27.50 &  0.63\\
  27 &     1.00 &     5.85 &     0.87&&&&-0.31 &   -23.44 &  0.64\\
  28 &     1.00 &    11.26 &    1.00&&&&-0.57  &   -29.84 &  0.63\\
  29 &     1.00 &     9.39 &     0.98&&&&-0.46 &   -28.14 &  0.67\\
  30 &     1.00 &    10.47 &     1.01&&&&-0.54 &   -26.45 &  0.56\\
\enddata
\end{deluxetable}

\end{document}